\documentclass[12pt]{article}
\usepackage{latexsym,graphicx,multirow}
\usepackage{float}
\usepackage{amssymb}
\usepackage{amsmath}
\usepackage{amscd}
\usepackage{amsthm}
\usepackage[left=2cm,top=2.5cm,right=2.5cm,bottom=1.5cm]{geometry}
\usepackage{hyperref}
\usepackage{epstopdf}
\usepackage{cite}

\begin{document}
	
	\begin{center}
	\large{\bf{ Barrow HDE model for Statefinder diagnostic in non-flat FRW universe }} \\
	\vspace{10mm}
	\normalsize{ Archana Dixit$^1$, Vinod Kumar Bharadwaj$^2$, Anirudh Pradhan$^3$  }\\
	\vspace{5mm}
	\normalsize{$^{1,2,3}$Department of Mathematics, Institute of Applied Sciences and Humanities, GLA University,
		Mathura-281 406, Uttar Pradesh, India}\\
	\vspace{2mm}
	$^1$E-mail: archana.dixit@gla.ac.in\\
          $^2$E-mail: dr.vinodbhardwaj@gmail.com\\
         $^3$E-mail: pradhan.anirudh@gmail.com \\
         
	\vspace{10mm}
 \end{center}
\begin{abstract}
In this work we study a non-flat Friedmann-Robertson-Walker universe filled with a pressure-less dark matter (DM) and Barrow holographic 
dark energy (BHDE) whose IR cutoff is the apparent horizon. Among various DE models, (BHDE) model shows the dynamical enthusiasm to 
discuss the universe's transition phase. According to the new research, the universe transitioned smoothly from a decelerated 
to an accelerated period of expansion in the recent past. We exhibit that the development of $q$ relies upon the type of spatial curvature.
Here we study the equation of state (EoS) parameter for the BHDE model to determine the cosmological evolution for  the non-flat universe. 
The (EoS) parameter and the deceleration parameter (DP) shows a satisfactory behaviour, it does not cross the  the phantom line. 
We also plot the statefinder diagram to characterize the properties of the BHDE model by taking distinct values of barrow exponent $\triangle$.
Moreover, we likewise noticed the BHDE model in the $(\omega_{D}-\omega_{D}^{'})$ plane, which can furnish us with a valuable, powerful 
finding to the mathematical determination of the statefinder. In the statefinder trajectory, this model was found to be able to reach the 
$\Lambda CDM$ fixed point. 

\end{abstract}
 
 
 
 \section{Introduction}
Cosmology is one of the most dynamic areas in research. As we know that modern cosmology was developed with the birth of 
``general relativity" (GR)\cite{ref1,ref2}. In this manner, GR plays a very important role in the study of ``gravity" and ``cosmology". 
However, GR is still brimming with potential outcomes that require further investigation. The nature of singularities and the fundamental 
properties of the ``Einstein field equations'' (EFEs) are being studied by mathematical relativists. After the establishing of modern 
cosmology with the approach of GR, researchers became interested in the causes of the universe accelerating expansion. The biggest 
challenge in modern cosmology is the late-time behavior of the universe. Latest observational confirmations from type ``Ia supernovae"  
point out an accelerating phase of the universe. The driving force behind this exponential expansion is driven by an unknown material 
known as  ``dark energy" (DE),  dubbed with modified gravity theories. In literature, numerous modified theories of gravity have been 
proposed and discuss the accelerated expansion of the universe, namely ``f(R) gravity \cite{ref3,ref4,ref5,ref6,ref7,ref8}, 
$f(T )$ gravity \cite{ref9}, $f(R, T)$ gravity \cite{ref10}-\cite{ref12}, $f(G)$ gravity \cite{ref13}, scalar-tensor theories 
\cite{ref14} etc.''\\

In 2004, Huang and Li \cite{ref15} introduced a new DE model, which depends on the holographic principle (HP). According to (HP), 
``the number of degrees of freedom in a bounded system should be finite and related to the area of its boundary".
In this context, Gao et al. \cite{ref16} worked on as HDE model from Ricci scalar curvature. In this manuscript, we propose to 
replace the ``future event horizon area" with the inverse of the`` (Ricci scalar curvature)". Majumdar and Chattopadhyay \cite{ref17} 
proposed the MHRDE model in the structure of ``f(T) modified gravity" and analyses the statefinder hierarchy. Malekjani et al. \cite{ref18} 
have worked on the ADE model in the non-flat universe and discussed the statefinder (SF) diagnostic and  $(\omega_{D}-\omega_{D}^{'})$ plane.\\

When exploring the dark energy  of the universe, several common questions are raised. For example, what is the  shape or curvature of our Universe? 
In the Friedmann equation, when we discuss our Universe's history, ``past, present, and future" curvature $k$ plays a very significant role. 
Here $k$ has three distinct prospects, which can either be ``closed, open or flat". An open Universe $(k<0)$ is assumed to be ``hyperbolic" 
and  it is infinite in all directions for the homogeneity and isotropy. A flat Universe $(k =0)$ has no curvature and it is also 
similar to the open Universe. A closed Universe $(k>0)$ has a ``spherical''  curvature and has a defined surface area, for the 
homogeneity and isotropy. In this direction, we found in literature, the spatially flat`` running vacuum model" (RVM) has been broadly explored 
\cite{ref19}-\cite{ref28}. In past \cite{ref29,ref30}, it has been proposed to solve the ``coincidence problem", where the ``cosmological constant" 
term is considered to be varying with the ``Hubble parameter $(H)$". In above mentioned models various DE models have been proposed to depicts the 
accelerated phase of the universe. The behaviour of DE in these models is highly model subordinate and separating between different (DE) 
by the sensitive diagnostic tool.\\

In this context, Sahni et al. \cite{ref31,ref32} introduced the geometrical statefinder diagnostic tool that can discriminate various DE models.
Using Statefinder diagnostic analysis, several authors have researched HDE models in non-flat universe  \cite{ref33}. From previously published 
works \cite{ref34,ref35,ref36,ref37,ref38,ref39} there is a clear juxtaposition with the different versions or other kinds of cut-off scales 
for the HDE model. Along with the statefinder diagnostic, the another analysis is $\omega_{D}-\omega_{D}^{'}$ that differentiate various DE models 
are generally mentioned in the literature. Motivated by the holographic fundamentals and utilizing different framework entropies, some new types 
of DE models were proposed such as, the ``Tsallis holographic dark energy" (THDE) \cite {ref40,ref41}, the ``Tsallis agegraphic dark energy" 
(TADE)  \cite{ref42} ``the Renyi holographic dark energy" (RHDE)  \cite{ref43,ref44}, and ``the Sharma-Mittal holographic dark energy" (SMHDE) 
model \cite{ref45}. Many authors have recently shown a great interest in HDE models and explored in different context \cite{ref46}-\cite{ref49}.\\

The researchers \cite{ref50,ref51} developed ``Barrow holographic dark energy". Here  the authors utilized the (HP) in 
a cosmological structure and Barrow entropy rather then the well-known ``Bekenstein-Hawking". Recently it was shown that ``quantum-gravitational" 
effects might introduce deformations on the black hole surface, which, although complex and dynamical. The dark energy EoS parameter was obtained by 
expressing the effect of barrow exponent  and showing distinct dark energy scenarios. Following this idea \cite{ref52,ref53} of Barrow entropy 
is considered as

 \begin{equation}
 \label{1}
 S_{B}= \left(\frac{B}{B_{0}}\right)^{1+\frac{\triangle}{2}}, 
 \end{equation}
where $B$ is the ``normal horizon area" and $B_{0}$ is the ``Planck area". The new ``barrow exponent" $\triangle$ is the 
``quantum-gravitational deformation" and is bound as $0 \le \triangle \le 1$ \cite{ref54}-\cite{ref58}, with $\triangle = 1$ comparing 
to the most complex and fractal structure, while $\triangle = 0$ relates to the easiest horizon structure in this case, the 
``standard Bekenstein-Hawking entropy" is reestablished. It is notable that for the special case $\triangle = 0$, the above connection 
gives the standard HDE, i.e., $\rho_{_{D}} \propto L^{-2}$. Subsequently, the BHDE is definitely a most broad structure in comparison to 
the standard HDE situation. We focus on the distinct values of $(\triangle>0)$. Here we consider that the expected  
``Hubble Horizon ($H^{-1}$)" as the IR cutoff $(L)$, where $C$ is the unknown parameter. We propose  the 
energy density of BHDE as 

\begin{equation}
\label{2}
\rho_{_{D}}= C H^{2-\triangle}
\end{equation}
The Barrow entropy was applied in the framework of ``gravity-thermodynamics" conjecture \cite{ref55,ref56,ref57}.  
`` First law of thermodynamics" can be applied on the universe apparent horizon. Although this construction is very efficient in describing the 
late-time universe, one should carefully examine whether the aforementioned extra terms are sufficiently small not to spoil the 
early-time behaviour, particularly, the Big Bang Nucleosynthesis (BBN) epoch \cite{ref53}.
In this model, we consider a ``spatially  non-flat", homogeneous and isotropic space time as the underlying
geometry of the universe. To explain the ongoing accelerated expansion of the DE model, the Hubble horizon has been considered a suitable IR cutoff. \\

The authors \cite{ref58,ref59} have focused on Barrow HDE models for Statefinder diagnostic and Statefinder hierarchy respectively in 
their recent papers. They \cite{ref58} have discussed BHDE model for flat FLRW universe. We have revisited in \cite{ref58} and obtained 
the BHDE model for non-flat FRW universe and analyzed the comparative difference between these two papers.
This work aims to  use the diagnostic tools to differentiate between the  BHDE models  with various esteem of $\triangle$. The 
plan of this manuscript  is as follow: In  section $2$, we introduce the Basic field equation for the proposed BHDE model. The Statefinder 
diagnostic is described in Sect. $3$. We explore the $\omega_{D}$-$\omega_{D}^{'}$ plane in Sect. $4$. Finally, conclusions and discussions
 are given in Sec. $5$.
 
\section {Basic field Equations} 

The metric of the ``Friedmann-Robertson-Walker Universe" is given by

\begin{equation}
\label{3}
ds^2=dt^2-a^{2}(t) \left(\frac{dr^2}{1-k r^2}+r^2 d\Omega^2\right),
\end{equation}
where $k$ represents the curvature of the space with $k = -1$, 0 and $1$ referring to open, flat and closed universes. The IR cut-off related to 
the universe in the holographic model is $ L=\frac{1}{H}$. Subsequently, to connect  the curvature of the universe to the energy density, 
we utilize the First ``Friedmann equation'' written as:
\begin{equation}
\label{4}
H^2+\frac{k}{a^2}=\frac{1}{3 M^{2}_{p}} \left(\rho_{m}+\rho_{D}\right)
\end{equation}
We likewise characterize the dimensionless density parameters $ \Omega_m=\frac{\rho_{m}}{3 M^{2}_{p} H^2} $, $ \Omega_D=\frac{\rho_{_D}}{3 M^{2}_{p} H^2}$ 
and $ \Omega_k=\frac{k}{a^2 H^2} $, where $\Omega_{m}$ is the non-relativistic matter-energy density parameter, $\Omega_{_{D}}$
is the  density parameter of DE, and $\Omega_{k}$ is the curvature density parameter. Therefore, we can rewrite the ``first Friedmann equation" as
($\Omega_m+\Omega_D=1+\Omega_k$). The ``conservation law"  for matter and  ``Barrow holographic dark energy"  defined as:
$\dot\rho_{m}+3H\rho_{m}=0, \dot\rho_{_{D}}+3H(p_{_{D}}+\rho_{_{D}})=0$. where $\omega_{_{D}} = p_{D}/\rho_{_D}$ is the EoS parameter of 
the BHDE, $\rho_{m}$ and $\rho_{_D}$ are the energy densities of DM and DE, respectively. It should be  emphasized that, in the non-flat
universe the characteristic length which plays the role of the IR-cutoff is the radius $L$ of the event horizon measured on the sphere 
of the horizon. In the recent work, the Barrow holographic dark energy as taking the time derivative of Eq. (2)
\begin{equation}
\label{5}
\dot{\rho_{D}}=(2-\Delta) \rho_{D} \frac{\dot{H}}{H}
\end{equation}
Combining the Eqs. (4) and (6)
\begin{equation}
\label{6}
\frac{\dot{H}}{H^2}=\frac{-3 \Omega_m +2 \Omega_k}{\Delta+(\Delta-2) \Omega_k+(2-\Delta) \Omega_m}
\end{equation}
The ``deceleration parameter" $(DP)$ for the BHDE model can be determined by

\begin{equation}
\label{7}
q=-1-\frac{\dot H}{H^{2}}.
\end{equation}

Following Eq. (6) DP is obtained as
\begin{equation}
\label{8}
q=\frac{(\Delta +1) \Omega _m-\Delta  \left(\Omega _k+1\right)}{\Delta +(\Delta -2) \Omega _k-(\Delta -2) \Omega _m}
\end{equation}
we defined the  EoS parameter by using the Eqs. (5)-(6) :
\begin{equation}
\label{9}
\omega_{D}=\frac{(\Delta -2) \Omega _k-3 \Delta }{3 \left(\Delta +(\Delta -2) \Omega _k-(\Delta -2) \Omega _m\right)}
\end{equation}

Similarly  we describe  $\omega_{_{D}}^{'}$ and $\Omega_{_{D}}^{'}$ by differentiating the $\omega_{_{D}}$ and
 $\Omega_{_{D}}$ with respect to $lna$ and  using the Eq. (9) , we find

\begin{equation}
\label{10}
\omega^{'}_{D}= -\frac{(\Delta -2) \left(-3 \Omega _k \Omega _m \left(2 (\Delta -4) \Delta +\left(\Delta ^2-4\right) 
\Omega _m\right)+\Omega _k^2 \left(\left(3 \Delta ^2-4 \Delta -4\right) \Omega _m-16 \Delta \right)+9 \Delta ^2 \left(\Omega _m-1\right) 
\Omega _m\right)}{3 \left(\Delta +(\Delta -2) \Omega _k-(\Delta -2) \Omega _m\right){}^3}
\end{equation}
\begin{equation}
\label{11}
\Omega^{'}_{D}=\frac{\Delta  \left(\Omega _k-\Omega _m+1\right) \left(3 \Omega _m-2 \Omega _k\right)}{\Delta +(\Delta -2) 
\Omega _k-(\Delta -2) \Omega _m}
\end{equation}
In FRW cosmology, ``the Ricci scalar $(R)$'' 
 is defined as ``$R=-6 \left(\dot{H}+2 H^2 +\frac{k}{a^2}\right)$"
.Here we consider a dark energy component, which is proportional to the inverse of squared ``Ricci scalar curvature". Our investigation 
is predominantly focused on the cosmological development of BHDE and examined with statefinder diagnostic.

 Utilizing the DE density as the BHDE in the Friedmann equation, we get

 \begin{equation}
 \label{12}
 H^{2}=\frac{1}{3}8\pi G  \rho_{_m0} e^{-3x}+(\alpha-1)k e^{-2x} +\frac{1}{2}\frac{dH^{2}}{dx}+2H^{2}
 \end{equation}

Here we  take $x= lna$ and solving the  Eq. (12), we get

\begin{equation}
   \label{13}
 E^{2}(a) = \Omega_{m0}a^{-3}+\Omega_{k0}a^{-2}+\frac{2}{2-\alpha}\Omega_{m0}a^{-3}+f_{0}a^{-{4-\frac{2}{\alpha}}},
 \end{equation} 
where $f_{0}$ is an integration constant,   $\Omega_{m{0}} = \rho_{_m0}/3M_{p}^{2}H^{2}$ is the current density parameter of non-relativistic 
matter and,  $E^{2} = \frac{H(z)^{2}}{H_{0}}$ is the reduced Hubble parameter, proposed on elective approach to use the IR cutoff by utilizing 
the Hubble scale. In this model, the Hubble IR cutoff gives a working model of BHDE. Here we assume scale factor as  $a = (1+z)^{-1}$, for 
solving the cosmological parameters. From Eq. (13) the value of $f_{0}$ is found to be $f_{0} = 1- \Omega_{k0}-\frac{2}{2-\alpha}\Omega_{m{0}}$. 
For the dark energy scenario, the value of $\alpha$ lies in range $1/2 <\alpha<1$. It is observed that this model is almost identical  to the Ricci-DE. 
Here the Ricci scalar is a linear combination of the Hubble parameter and its time derivative in flat spatial parts without spatial curvature 
\cite{ref54}.
  
\begin{figure}[H]
 	\centering
 	(a)\includegraphics[width=7cm,height=7cm,angle=0]{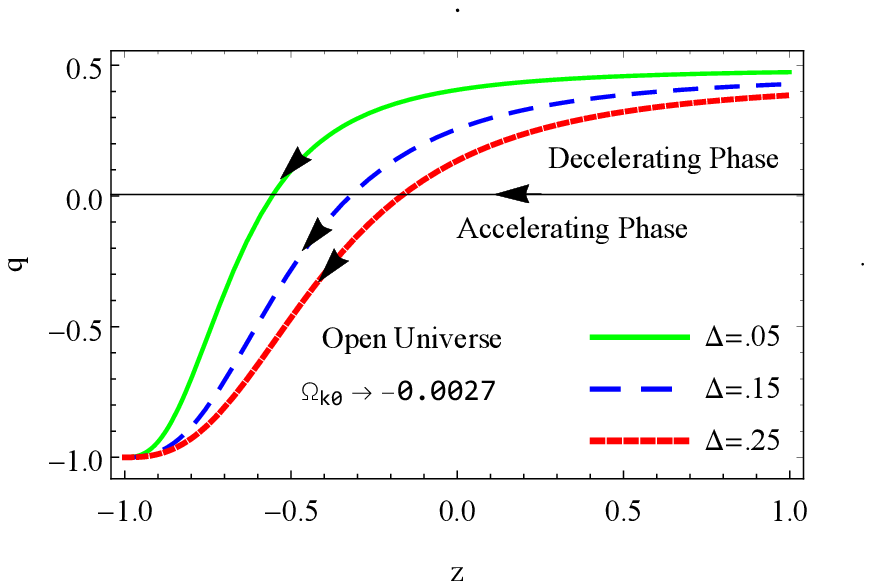}
 	(b)\includegraphics[width=7cm,height=7cm,angle=0]{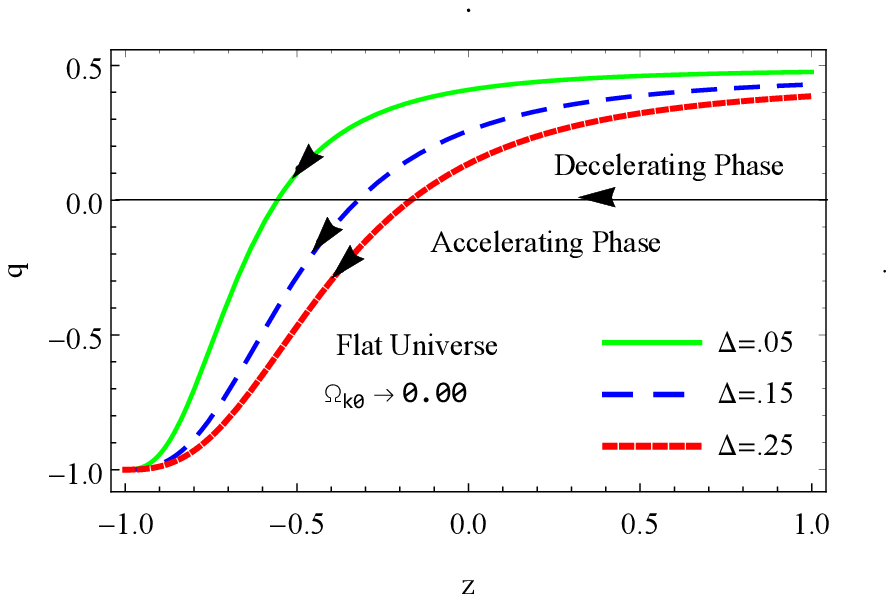}
 	(c)\includegraphics[width=7cm,height=7cm,angle=0]{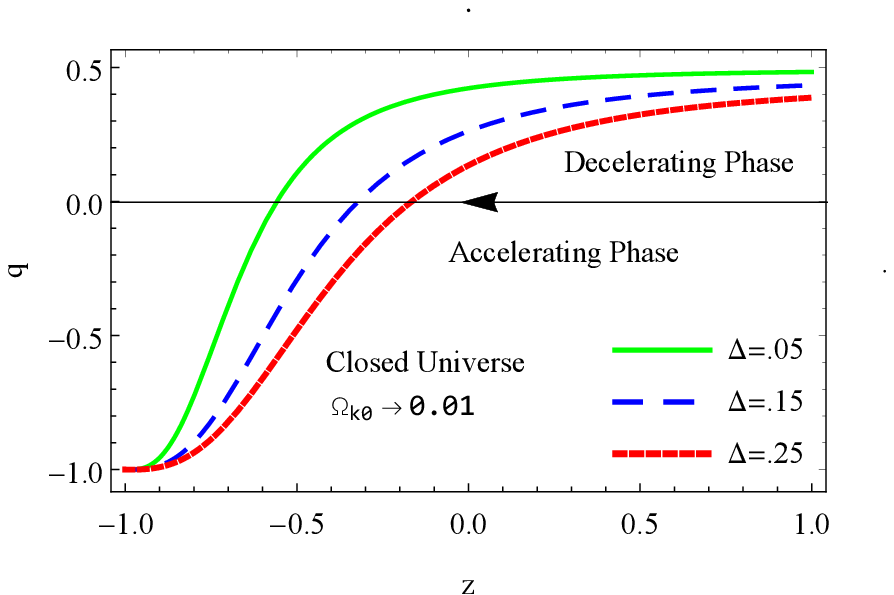}
 	\caption{ Evolution of DP (q) for BHDE model versus redshift $(z)$ for various esteem of $\triangle$ by taking $\Omega_{k0}=-0.0027$ 
 	for open, $\Omega_{k0}=0.00$ for flat and $\Omega_{k0}=0.01$ for closed. }
 	 \end{figure}
   Figures 1(a)$-$1(c) show the plots  for the evolution of $q$. As we noted in the figure, for a non-flat BHDE model, the deceleration parameter 
   is a redshift ($z$) function. The transition redshift depends on the possible values of the Barrow exponent $\triangle$. 
   Our graphs are plotted for $\Omega_{k0}= - 0.0027, ~ 0.00$ ~and ~ $0.01$ corresponding to  open, flat and, closed universes, respectively, in the 
   light of the  ``Planck 2018  cosmological observational data". We noticed that the (DP) demonstrate a universe with an accelerating expansion 
   rate, and it can be also observed that BHDE model is entering the accelerating phase at the redshift $-0.7< z_{t}<0$. Moreover, we analyze 
   at near the high redshift 
   region, we have ``$q\to-1$, while at $z\to-1$" for the open, flat and closed universe. It should be noted that for $z<-1$ the cosmos will cross 
   the phantom $(q <-1)$, for distinct values of $\triangle$. For $\triangle = 0.05, 0.15,0.025$, universe transit from deceleration to 
   acceleration at late time.\\

 \begin{figure}[H]
  	\centering
  	(a)\includegraphics[width=7cm,height=7cm,angle=0]{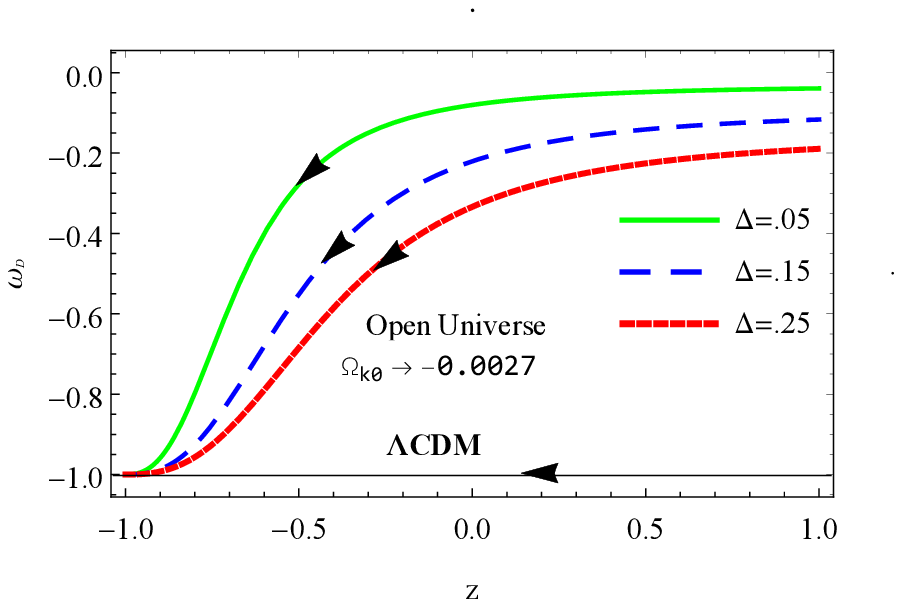}
  	(b)\includegraphics[width=7cm,height=7cm,angle=0]{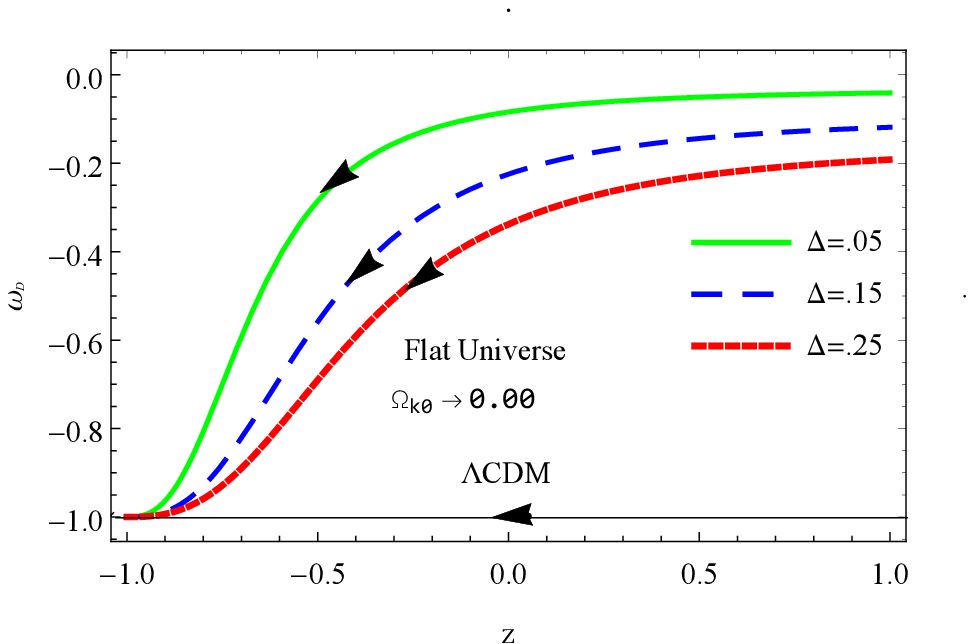}
  	(c)\includegraphics[width=7cm,height=7cm,angle=0]{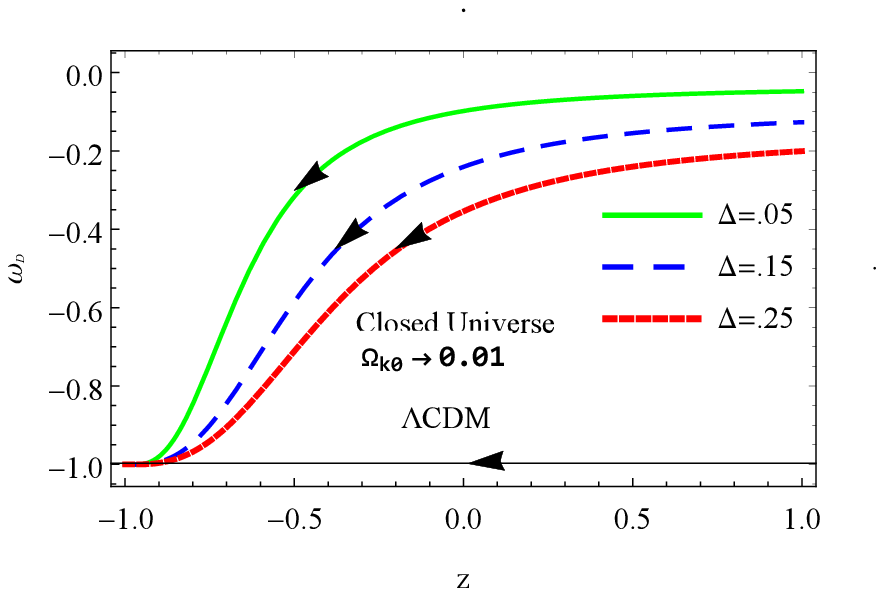}
    \caption{ Estimation of EoS $(\omega)$ for BHDE model versus redshift $(z)$ corresponds to various esteem of $\triangle$ by 
    taking $\Omega_{k0}= - 0.0027$ for open, $\Omega_{k0}=0.00$ for flat and $\Omega_{k0}=0.01$  for closed. }
\end{figure}

The estimation of Eos parameter for DE is the major effort in observational cosmology. In figures 2(a)$-$2(c), we depict the behaviour of 
EoS parameter versus $z$ we noticed that at high redshifts, the EoS parameter is nearly zero for the open, flat and closed universe, so the 
{ BHDE} behaves just like dark matter. The EoS( $\omega$) approaches $-1$ at $z\to -1$ show the consistency of the model, with `` 
$\Lambda$ CDM model". In this model the EoS parameter  lies  in  the  region $ -1 \leq \omega_{_{D}} < 0$, which is good agreement with 
the accelerating universe. Also, we observe in Figs. 2(a)$-$2(c) for open, flat, and close universe, the EoS parameter does not cross  
the  cosmological constant boundary $``\omega=-1" $ for various estimation of barrow exponent. Also we conclude that our model represents an HDE model 
for $\triangle=2$ and its shows a cosmological constant $\omega =-1$. Interestingly, we noticed that for various esteems of 
``$\triangle = 0.05, 0.15, 0.25$" EoS parameter lies in quintessence region. Here the value of $\triangle$ decreases the profile 
shift towards higher values at redshift $z =0$ and beyond. Finally, in the 
far future, we can mathematically measure the asymptotic value of $\omega_{D}$. This means that  the dark energy EoS parameter has an 
interesting behaviour in $\Lambda CDM$ cosmology.


\section{Statefinder}

Statefinder (SF) parameters are the diagnostic and sensitive tool, which is used to segregate between different DE models. 
The degeneracy of these parameters, the Hubble parameter $H$ and the DP $(q)$ does not differentiate between different DE models. 
Sahni et al. \cite{ref31} presents a set of parameters ${(r,~s)}$  called the statefinder, characterized as,

 \begin{equation}
 \label{14}
 r=1+\Omega_k+\frac{9}{2}  \omega_{d}  (\omega_{d}+1) \Omega _d-\frac{3 }{2} \omega^{'}_d \Omega _d 
 \end{equation}
 
 \begin{equation}
 \label{15}
 s=1+\omega_{d}-\frac{1}{3}\frac{\omega^{'}_d}{\omega_{d}}
 \end{equation}
 
  The $(r, ~s)$ plot of DE models can assist with separating and segregate different models. In the observed $\Lambda CDM$ model, 
  the ($r, ~s$) direction is relates to fixed point, with $(r = 1)$ and $(s = 0)$ \cite{ref31}.  In the literature, the cosmological behaviour 
  of different DE models, including BHDE, RHDE and HDE was examined and differentiated with statefinder parameters 
  \cite{ref61}-\cite{ref64}.
 
\begin{figure}[H]
	\centering
	(a)\includegraphics[width=7cm,height=7cm,angle=0]{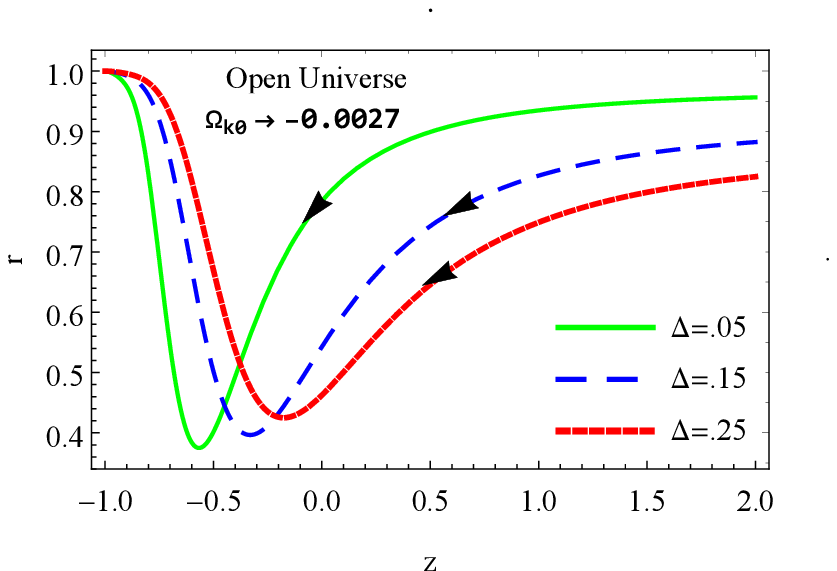}
	(b)\includegraphics[width=7cm,height=7cm,angle=0]{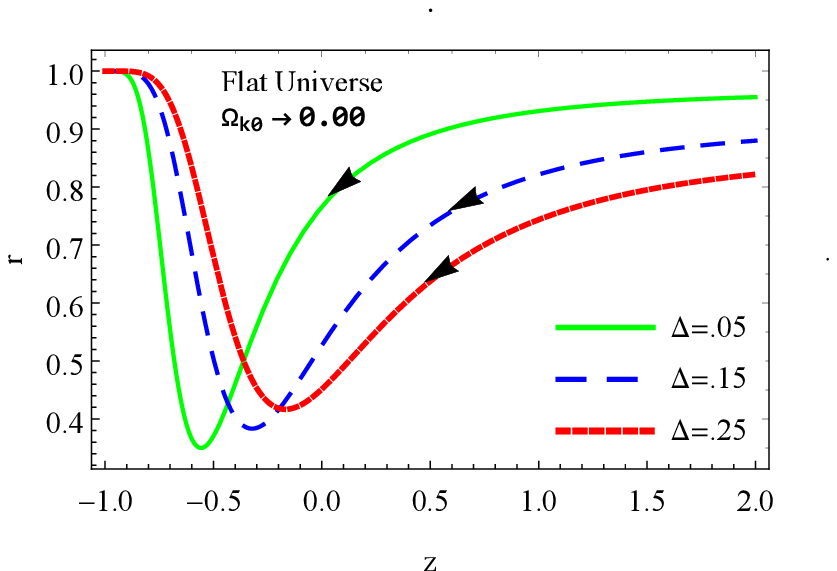}
	(c)\includegraphics[width=7cm,height=7cm,angle=0]{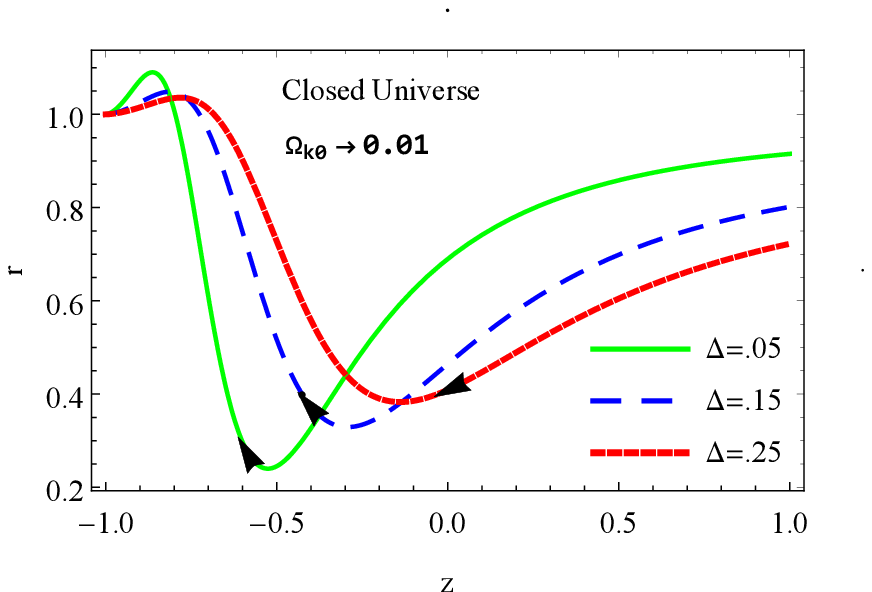}
	\caption{ Evolution of SF (r) for BHDE model versus redshift $(z)$ for various esteem of $\triangle$ by taking 
	$\Omega_{k0}= - 0.0027$ for open, $\Omega_{k0}=0.00$ for flat and $\Omega_{k0}=0.01$ for closed. }
\end{figure}

The evolution of  $(r, ~s)$ against redshift $(z)$ for the FLRW non-flat universe has been explored in Figs. (3a)$-$(3c) for the various 
spatial curvatures of the universe. The first parameter $(r)$ of ``Oscillating dark energy" (ODE), near the high redshift, approaches 
to standard `` $\Lambda$CDM" behaviour. At the same time it deviates entirely from the normal behaviour at low redshift, the second parameter 
$(s)$  in  Figs. (4a)$-$(4b) depicts the opposite in behavior \cite{ref65,ref66}. In this context Figures (3a$-$3b) $\&$ (4a-4b) shows the 
assessment for various estimations of Barrow parameter $\triangle =0.05, 0.15,0.25 $ and approaches
to the $\Lambda$ CDM, by taking the value (for $\Omega_{m_{0}}$ = 0.27 and$ H_{0}$= 69.5) for open, flat and closed universe.
As predicted, the above adjusted Friedmann equations are reduced to the $\Lambda$CDM scenario for $\triangle= 0$.

 \begin{figure}[H]
 	\centering
 	(a)\includegraphics[width=7cm,height=7cm,angle=0]{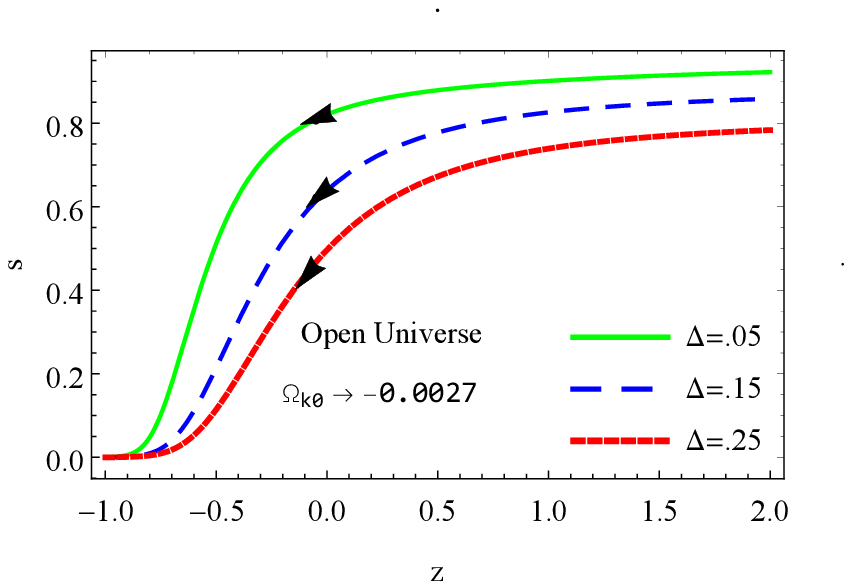}
 	(b)\includegraphics[width=7cm,height=7cm,angle=0]{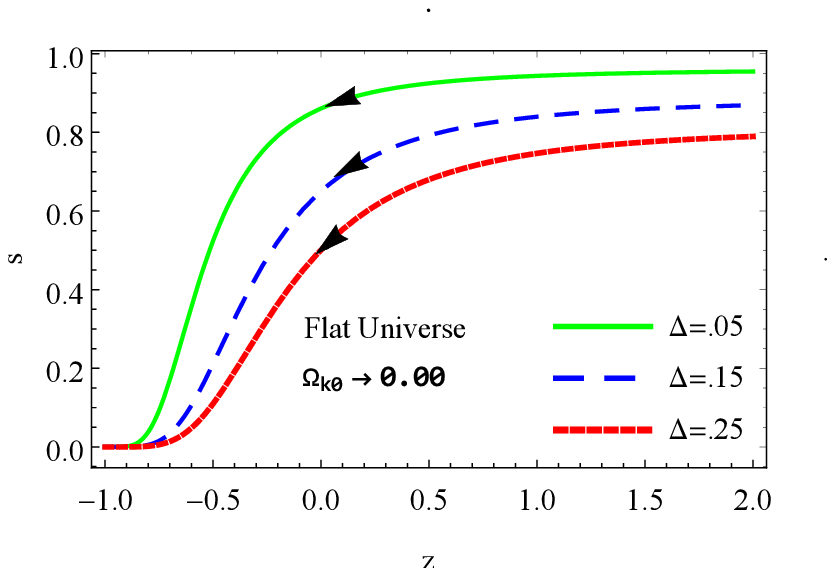}
 	(c)\includegraphics[width=7cm,height=7cm,angle=0]{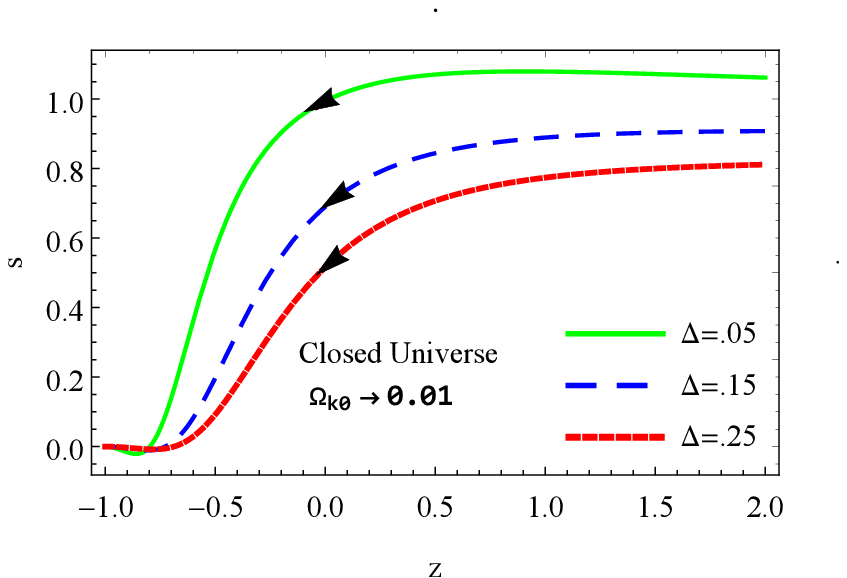}
 	
 	\caption{ Evolution of statefinder (s) for BHDE model against redshift $(z)$ for various esteem of $\triangle$ by 
 	taking $\Omega_{k0}= - 0.0027$ for open, $\Omega_{k0}=0.00$ for flat and $\Omega_{k0}=0.01$ for closed. }
 	 \end{figure}
  \begin{figure}[H]
  	\centering
  	(a)\includegraphics[width=7cm,height=7cm,angle=0]{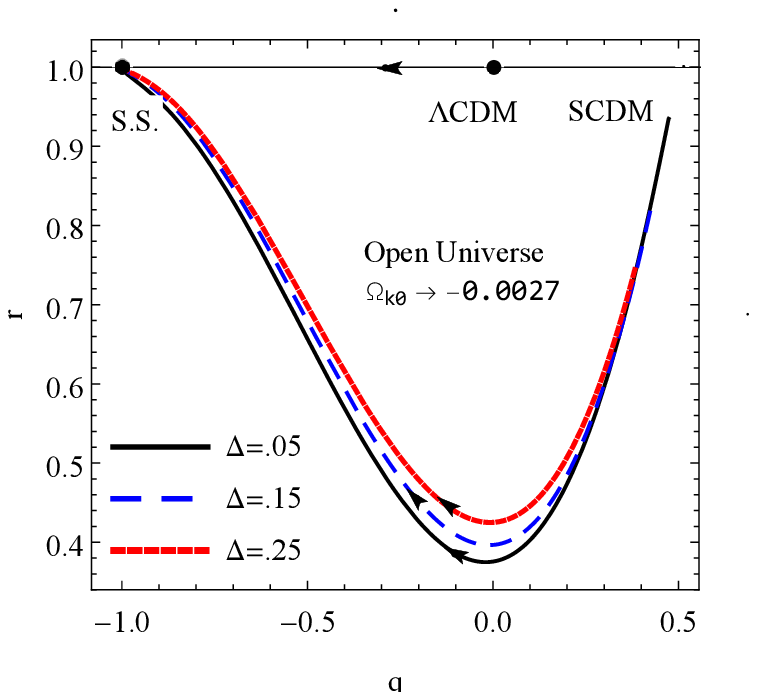}
  	(b)\includegraphics[width=7cm,height=7cm,angle=0]{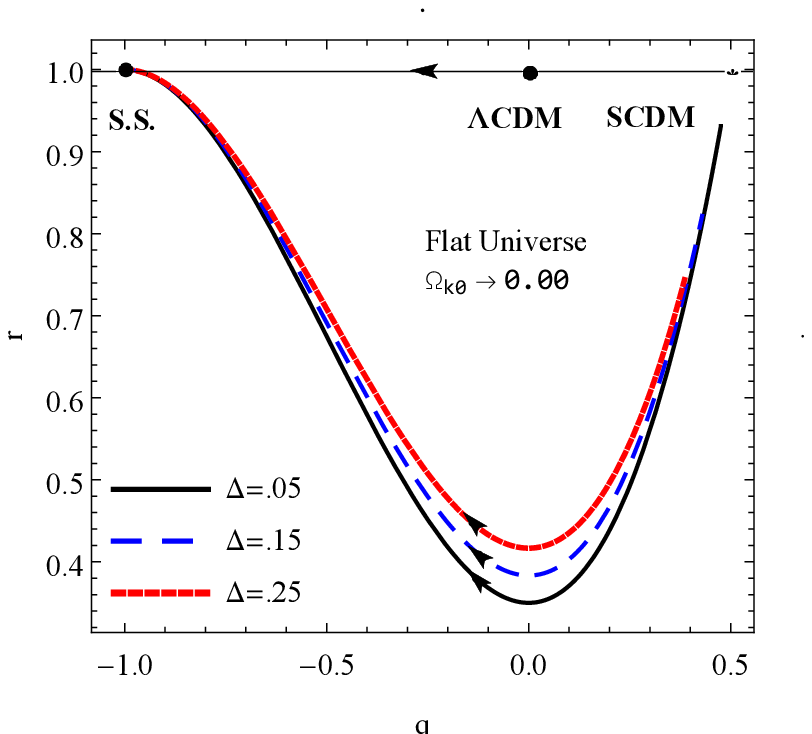}
  	(c)\includegraphics[width=7cm,height=7cm,angle=0]{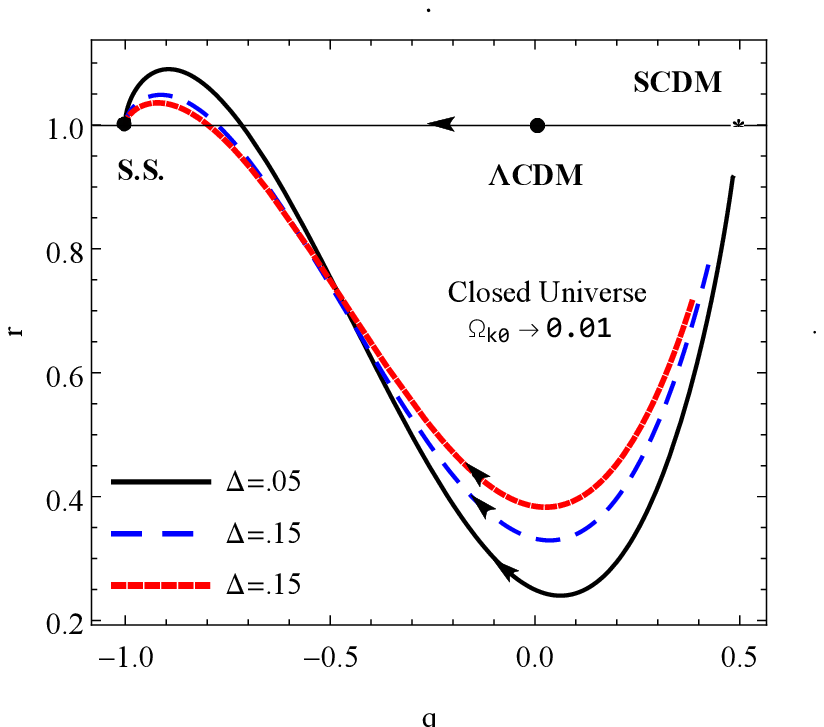} 
  	\caption{ Evolution of trajectories $(r-q)$ for BHDE model against redshift $z$ for various esteem of $\triangle$ 
  	by taking $\Omega_{k0}= - 0.0027$ for open, $\Omega_{k0}=0.00$ flat and $\Omega_{k0}=0.01$ closed. }
  \end{figure}

As a result, Figs. 5(a)$-$5(c) indicate the evolutionary trajectory in $(r-q)$ plane.
The horizontal line at ``$r = 1$" compares to the time development of the $\Lambda$CDM model. 
The curves move from (+ve to -ve)  in $q$  clarifies the phase transition of the universe. The BHDE model may begin from the region of the 
``SCDM model ${(r,q)}$ = ${(1, 0.5)}$" for various esteem  of $\triangle = 0.05, 0.15, 0.25$. However, BHDE model approached to the $(SS$) 
model as $\Lambda$CDM and started from the SCDM model. The stars represent the $SCDM$, and dot represent Steady State $(SS)$ models, respectively. 
These figures demonstrate that BHDE model switches from deceleration to acceleration. The $(q-factor)$ still has adverse values,
starting from $q <-1$ and later tending to $q = -1$ with large values $\triangle$ see in table-1. From the above analysis, we conclude that our 
model converges to both regions (SS) model in late time as the $\Lambda CDM$ model  and started from SCDM for the different values of the 
``barrow exponent" for the open, flat and closed universe.


\begin{figure}[H]
	\centering
	(a)\includegraphics[width=7cm,height=7cm,angle=0]{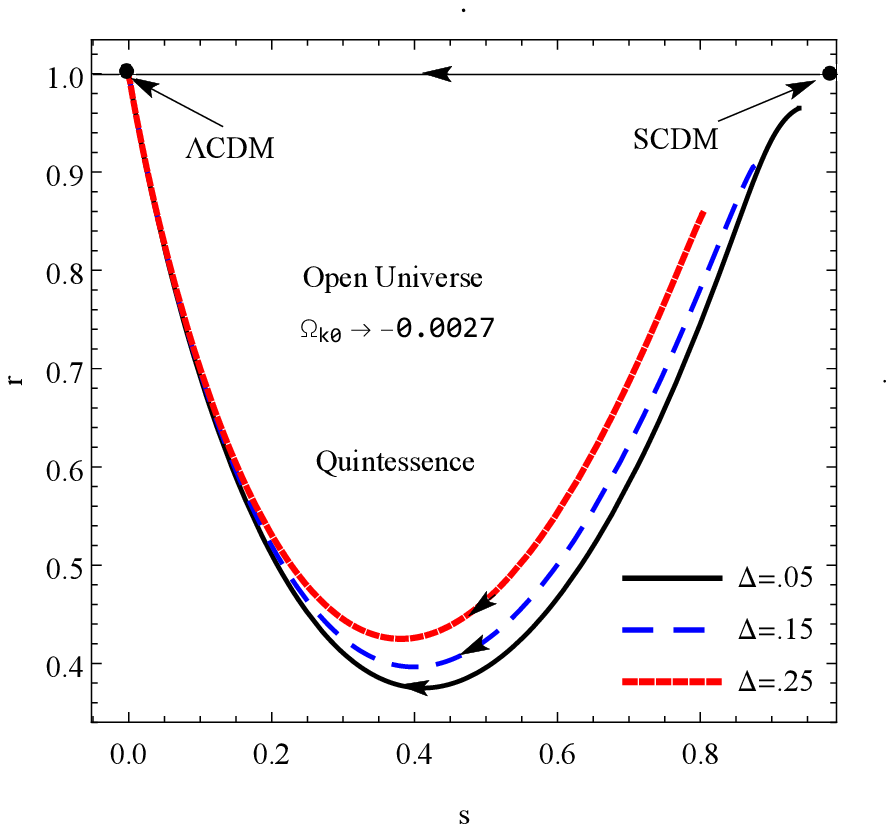}
	(b)\includegraphics[width=7cm,height=7cm,angle=0]{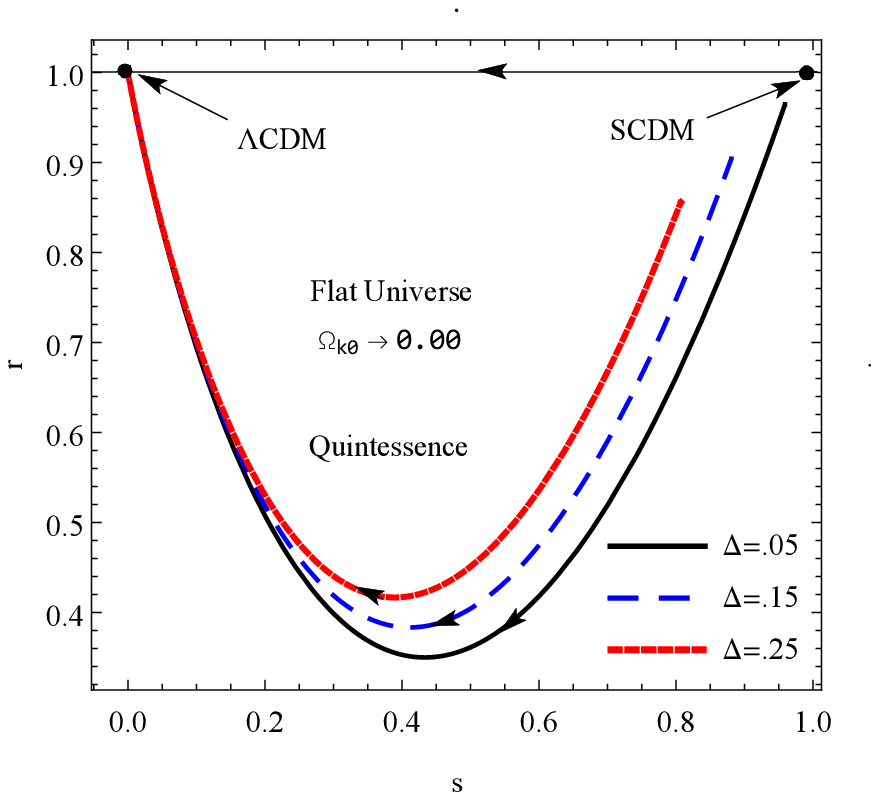}
	(c)\includegraphics[width=7cm,height=7cm,angle=0]{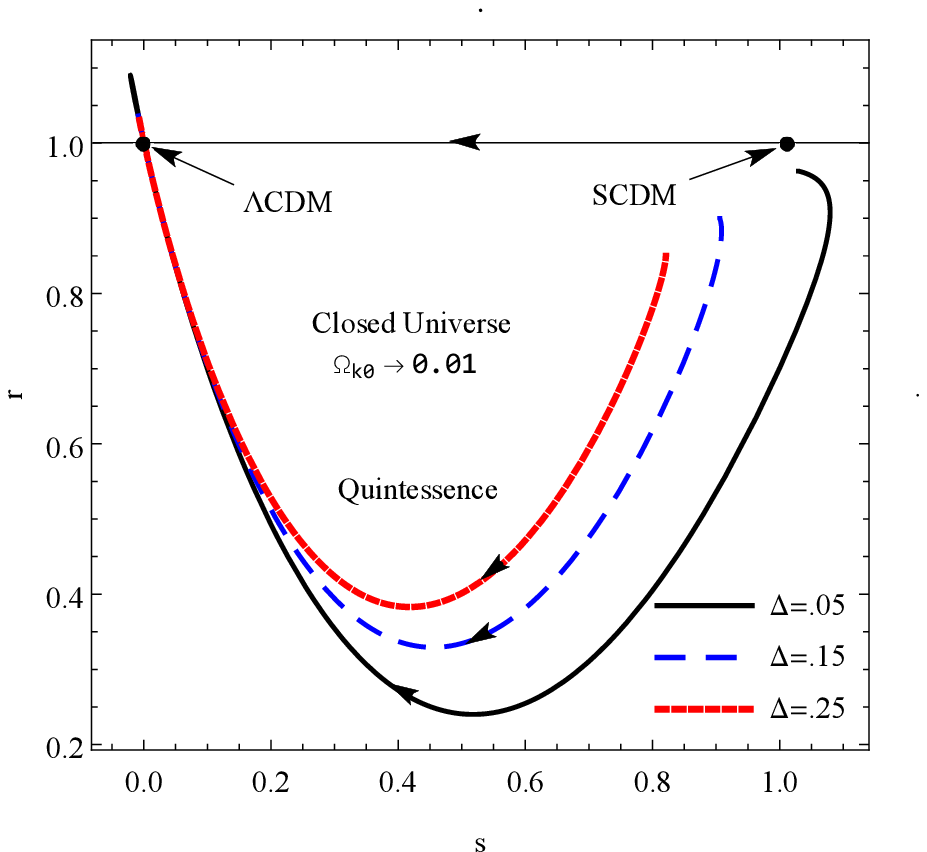}
	\caption{ Evolution of trajectories $(r-s)$ for BHDE model against redshift $(z)$ for various esteem of $\triangle$ 
	by taking $\Omega_{k0}= - 0.0027$ for open, $\Omega_{k0}=0.00$, for flat and $\Omega_{k0}=0.01$ for closed. }
\end{figure}

The evolution trajectories in $(r-s)$ plane are shown in Figs. 6(a)$-$6(c), for various Barrow exponent parameters 
$\triangle$. In these figures, we are using distinct spatial curvatures like $\Omega_{k0} = 0.00, - 0.0027$, and $0.01$ for the flat, open and 
closed universes respectively. As the universe is expanding, the statefinder begins at an early stage from a fixed point $({r = 1, s = 0})$. 
As we see in figures, the $s$ parameter increases, and the $r$ parameter decreases. The arrows give the direction of evolution in the 
figures 6(a)$-$ 6(c). In other words, the trajectory begins in the $(Quientessence-region)$ during an early time and approaches to $\Lambda CDM$ 
in late time for the different approximation of $\triangle = 0.05,0.15,0.25$. In addition, the evolutionary trajectory is based 
on the curvature of the model of the universe. While the region $r < 1$, $s >0$, indicates a behaviour similar to  Quintessence $(q-model)$ 
\cite{ref31,ref32}. We may also observe that the BHDE model for the open flat and closed universe would approach the $\Lambda$CDM model in 
the future, (see table 1). In the late universe, we discovered that our BHDE model creates curved trajectories that are similar to the 
$\Lambda$CDM model.


\begin{table}[ht]
\caption{The values cosmological parameter $q$, $\omega$, $r$, $s$ and  $\omega_{D}$ for the open, flat and closed universe  }
		\centering
	\begin{tabular}{c c c c c c c}
		\hline\hline
		$Universe $ &$\triangle$ & $q$ & $r$ & $s$ & $\omega_{D}$ & $\omega_{D}^{'}$ \\
		[0.5ex]
		\hline

	open~~~ ($\Omega_{k0} = - 0.0027$)&	$0.05$    & $-0.538862$ & $0.712814$ & $0.0921477$ & $0.0921477$& $-0.436106$  \\
		
			~~~&$0.15$    & $ -0.806663$ & $0.939761$ & $ 0.0153671$ & $-0.880775$& $-0.274426$  \\
			~~~&$0.25$    & $-0.877692$ & $0.975956$ & $0.0921477$ & $-0.928654$& $ -0.182561$  \\
		
		\hline
		flat~~~ ($\Omega_{k0} = 0$) &$0.05$    & $-0.533623$ & $ 0.68732$ & $0.101831$ & $-0.689255$& $-0.431984$  \\
	
	~~~&$0.15$    & $-0.803247$ & $0.918655$ & $  0.020773$ & $-0.875627$& $-0.272144$  \\
	~~~&$0.25$    & $-0.875325$ & $ 0.961574$ & $0.0092306$ & $-0.92545$& $ -0.181351$  \\
	
		\hline
		closed~~~($\Omega_{k0} = 0.01$) &	$0.05$    & $-0.533623$ & $ 0.68732$ & $0.101831$ & $-0.689255$& $-0.431984$  \\
		
	~~~&	$0.15$    & $-0.803247$ & $0.918655$ & $  0.020773$ & $-0.875627$& $-0.272144$  \\
	~~~&	$0.25$    & $-0.875325$ & $ 0.961574$ & $0.0092306$ & $-0.92545$& $ -0.181351$  \\
			\hline
	\end{tabular}
	\label{table:1}
\end{table}


 \section{$\omega_{D}-\omega_{D}^{'}$ plane}
 
 The ($\omega_{D}-\omega_{D}^{'}$) plane, address as the dynamic property of the BHDE model. Many researchers \cite{ref67}-\cite{ref70} 
 recently investigated the emerging behavior of DE models of quintessence and tested the limits of quintessence in ($\omega_{D}-\omega_{D}^{'}$) 
 plane. The study of $\omega_{D}-\omega_{D}^{'}$ certainly gives us an alternate way to characterize DE models. Obviously, the 
 $\omega_{D}-\omega_{D}^{'}$ pair is connected to the statefinder pair ($r,s$). Figs. (7a)$-$(7c) gives an illustrative example of the evolution 
 of the BHDE in the $(\omega_{D}-\omega_{D}^{'})$ plane by fixing $\Omega_{_{D0}} = 0.73$ and varying  $\Omega_{_{k0}}$ = 0.0, - 0.0027, 0.01 
 corresponding to the flat, open and closed universe respectively. As shown in this figure, the value of $\omega_{D}$ decreases monotonically 
 while the value of $\omega_{D}^{'}$ first decreases from zero for all three cases open flat and closed. The curves  correspond to 
 $\triangle$ = 0.05, 0.15, 0.25 for the inclusion of different instances. The arrow denotes the  evolution of the direction. We see 
 clearly that the $\triangle$ parameter plays a key role in the model. 
As we know that the EoS parameter, describes the universe into various eras like ``radiation $\omega_{D} = -1/3$, matter $\omega_{D} = 0$ 
 and DE dominated $\omega_{D} = -1$." The DE era also divided into two regions  quintessence $(-1 < \omega_{D} \leq -1/3)$ and phantom 
 $(\omega_{D} < -1)$.\\
 
There are two distinct evolutionary zones, thawing and freezing, according to this plane. We observe from figure (7a$-$7c) 
the EoS parameter as well as its evolution measured by $\omega_{D}^{'}$ lie in the negative region. Observational evidences show that 
in the freezing region ($\omega_{D}^{'}<0$ with $\omega_{D}<0)$, the cosmos  expands with a more accelerated rate as compared to the thawing 
region ($\omega_{D}^{'}>0$ with $\omega_{D}<0)$. We also noticed for various esteem of Barrow exponent, our models lie in freezing region 
in the open, flat and closed universes. The evolutionary trajectory is also dependent on the $\triangle$ in seen Table 1. 
 
 
 \begin{figure}[H]
 	\centering
 	(a)\includegraphics[width=7cm,height=7cm,angle=0]{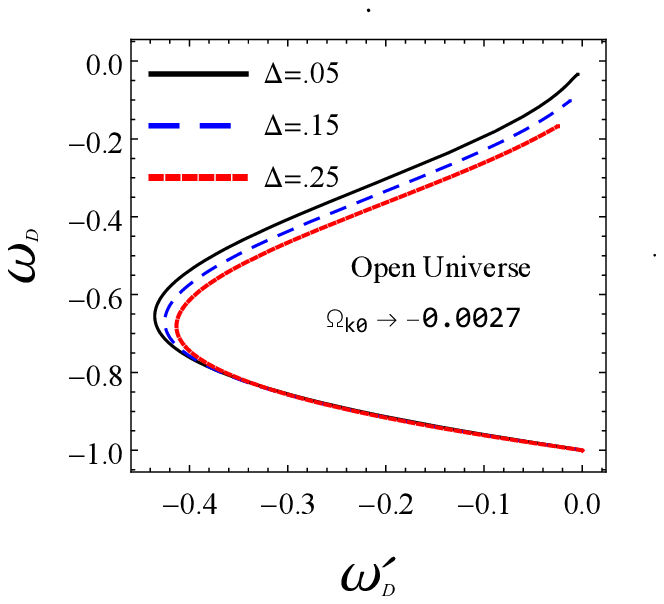}
 	(b)\includegraphics[width=7cm,height=7cm,angle=0]{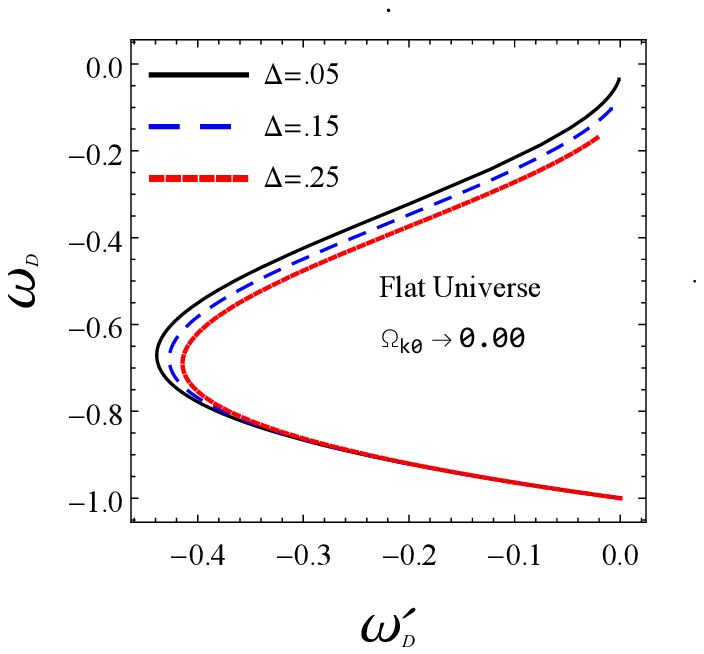}
 	(c)\includegraphics[width=7cm,height=7cm,angle=0]{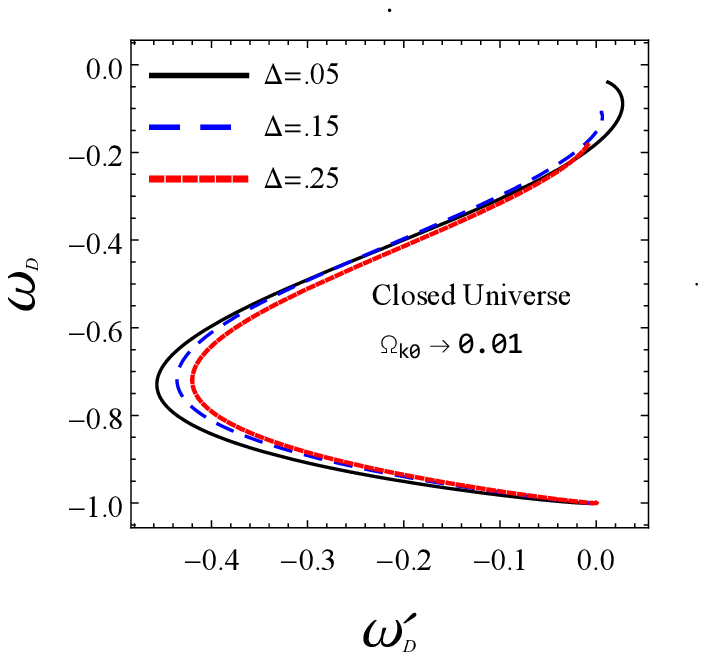}
 	\caption{ Evolution of  $(\omega_{D}-\omega_{_{D}})$ for BHDE model against redshift $(z)$ for various esteem of $\triangle$ 
 	by taking $\Omega_{k0}= - 0.0027$ for open, $\Omega_{k0}=0.00$ for flat and $\Omega_{k0}=0.01$ for closed. }
 \end{figure}
  \section{Conclusion}
 The Barrow entropy has been used to discuss the BHDE in this manuscript, which was involved in the usual Bekenstein-Hawking.  Our main 
  focus to diagnose BHDE model by the State finder  and $\omega_{D}-\omega_{D}^{'}$ plane.
  The  diagnostic tools which frequently apply to test the DE models are (a) $(r-s)$ parameter (b) $\omega_{D}-\omega_{D}^{'}$ plane. 
So, the major concern of the work is to apply diagnostic tools for the BHDE models.  \\
 
 The following are some of the models' main strengths: \\
 
  \begin{itemize}
  \item 
For the different values of the barrow exponent in Figs.1(a)$-$1(c), we first investigated the existence of our model's deceleration parameter.
We can see that the nature of the expansion is transit, that is, it is moving from deceleration (in the past) to acceleration (in the present). 
  
 \item 
The transition from the early deceleration period $(q > 0)$ to the present accelerating period $(q < 0)$ of the BHDE has been observed to be smooth.
The value of this redshift transformation is well matched with the existing cosmological findings, which is strong in agreement with the current findings.
  
 \item  
  The expression of the EoS parameter can be seen in the proposed BHDE model by changing the Barrow exponent $\triangle$ in Fig (2a$-$2c).
  For different parameter values, the EoS parameter of the BHDE model lies  in the quintessences region. Finally, in the far future, 
  we can analytically measure the asymptotic value of $\omega_{D}$. 
  
  \item 
  We also addressed the statefinder $(r- s)$ for various esteem of Barrow exponent $\triangle$. The behaviour  $r$ verses $z$ have 
  shown in figures (3a$-$3c) for the open, flat and close spatial curvatures. The $r(z$) oscillating dark energy (ODE) parameters approach 
  to the standard $\Lambda$CDM in the high redshift region. Similarly, $s(z)$ in Figs. (4a$-$4c), where $s(z)$ has the opposite behaviour as $r$. 
  
 \item  
  The excellent diagnostics of DE is represented by Figs. $5a$ and $5b$, which are $(r-s)$ and $(r-q)$ respectively for open flat and closed universe. 
  Here we take the value $\Omega_{m0}= 0.27$, $H_{0}= 69.5$ and  using the different values of $(\triangle = 0.05,0.15, 0.25)$, then the  
  averaged-over-redshift statefinder pair ($r- s$) obtained the steady state $(SS)$ model, quintessences ($q$-model)  and $(r- q)$ obtained  
  $(SS)$ model $\Lambda$CDM and SCDM. With BHDE, we can see that statefinders play a vital role in the FLRW universe. 
  Therefore the BHDE in FLRW  model gives more general results in comparison to $\Lambda$ CDM and $q$-model.
  
 \item  
 In the non-flat universe, we also performed the $\omega_{D}-\omega_{D}^{'}$ study for the interacting BHDE models in Figs. 7(a)$-$7 (c). 
 The $\omega_{D}-\omega_{D}^{'}$ analysis are useful method of dark energy. The  $\omega_{D}-\omega_{D}^{'}$ trajectories indicate the 
 freezing region for the BHDE model with Hubble horizon cut-off as seen in table-1.
 
  \item
  In the flat universe \cite{ref58}, the authors have shown that DE EoS parameter lies in the quintessence regime, phantom regime and also 
  cross the phantom-divide line during the cosmic evolution. But in the present study of non-flat universe, EoS parameter was found to lie in the region 
  $-1 \leq \omega_{D} < 0$ which is a good agreement with the accelerating universe. In our case, the EoS parameter does not cross $\omega = -1$.
  
  \item
  In the flat universe \cite{ref58}, the authors found the statefinder ($r-s$) trajectories divided into two regions, Chaplygin gas and 
  quintessence. But in our non-flat model, we find the statefinder ($r-s$) trajectories only lies in quintessence region.
    
\end{itemize}
 In summary, the $\Lambda$CDM model, which has an EoS $\omega = - 1$, is the simplest DE model. So far, this model has been regarded 
 as the mainstream model of cosmology due to its superior success in fitting current observational data.  
 We can conclude that the BHDE, model can be easily distinguished by using these diagnostic tools. Our analysis backs up the viability 
 of the BHDE with an IR cut-off, and cumulative observational data points to a future horizon.



\end{document}